\documentclass[10pt]{article}
\usepackage{fullpage}
\usepackage{amsmath}
\usepackage{amssymb}
\usepackage{amsfonts}
\usepackage{comment}
\usepackage[dvips]{epsfig}
\usepackage{color}
\usepackage{cite}

\def\##1{{\bf #1}}

\def\.{\mbox{ \tiny{$^\bullet$} }}
\def\,{\thinspace}

\def\c#1{\cite{#1}}

\def\r#1{(\ref{#1})}

\def\epso{\epsilon_{\scriptscriptstyle 0}}

\def\muo{\mu_{\scriptscriptstyle 0}}
\def\ko{k_{\scriptscriptstyle 0}}

\def\eps{\epsilon}

\date{}

\begin{document}

\baselineskip .582cm   

\vskip 0.2cm
\noindent {\large\bf BRUGGEMAN APPROACH FOR }\\
{\large\bf ISOTROPIC CHIRAL MIXTURES REVISITED}\\[20pt]
 {\bf Benjamin M. Ross} and {\bf Akhlesh Lakhtakia}
\\[20pt]
{\em CATMAS~---~Computational \& Theoretical Materials Sciences Group\\
Department of Engineering Science and Mechanics\\
Pennsylvania State University, University Park, PA 16802--6812, USA}\\[25pt]

\noindent {\bf ABSTRACT:}
Two interpretations of the Bruggeman approach for the homogenization of isotropic chiral  mixtures
are   shown to lead to different results. Whereas the standard interpretation is shown to yield the {\em average polarizability density\/} approach,
a recent interpretation turns out to deliver a {\em null excess polarization\/} approach. The difference between the two interpretations
arises from differing treatments of the local field.
\\


\section{Introduction}
Homogenization of particulate composite materials is at least a two--century--old theoretical problem;
yet, it retains its freshness to this day. Indeed, it can be argued that, as theoretical approaches can~---~at  best~---~only
{\em estimate\/} the  effective  constitutive parameters of a mixture of two or more component materials but still viewed as being
homogeneous, homogenization is unlikely to lose its charm for theorists for the foreseeable future \c{Mbook,Mackay}.

This viewpoint rose to the fore recently when we had occasion to look at the {\em Bruggeman approach\/} for
the homogenization of an isotropic mixture of two isotropic chiral materials. In this approach, the volume fractions
of both component materials are taken into account, but the particulate dimensions are effectively null--valued,
as explicated by Kampia and Lakhtakia \c{KL}.
In the {\em extended Bruggeman approach\/}, the particulate dimensions are considered as electrically small
but finite, as exemplified by Shanker \c{Shan}
for the chosen mixture. However, we found that Shanker's interpretation
of the Bruggeman approach differs in an essential point from that of Kampia
and Lakhtakia~---~in addition to the differing treatments of the particulate dimensions. Our ruminations on the
newly discovered difference led to this communication.

\section{Theory in Brief}
Let us consider an isotropic mixture of two isotropic chiral materials labeled $a$ and $b$. Their frequency--domain constitutive relations
are stated as
\begin{equation}
\left.\begin{array}{l}
{\bf D} = \epso\eps_p\, [{\bf E} + \beta_p \,\nabla\times {\bf E}]\\[2pt]
{\bf B} = \muo\mu_p\, [{\bf H} + \beta_p\, \nabla\times {\bf H}]
\end{array}\right\}\,,\quad (p = a,b)\,,
\label{eq1}
\end{equation}
where $\epso$ and $\muo$ are the permittivity and the
permeability of free space (i.e., vacuum); $\eps_{a,b}$ are the relative permittivity scalars, $\mu_{a,b}$ are the relative permeability scalars, 
and $\beta_{a,b}$ are  the chirality pseudoscalars
in the Drude--Born--Fedorov representation \c{LBel}; and an $\exp(-j\omega t)$ time--dependence is implicit. The volumetric
fractions of the two component materials are denoted by $f_a$ and $f_b=1-f_a$. The aim of
any homogenization exercise is to predict the quantities $\eps_{HCM}$, $\mu_{HCM}$ and $\beta_{HCM}$
appearing in the constitutive relations
\begin{equation}
\left.\begin{array}{l}
{\bf D} = \epso\eps_{HCM}\, [{\bf E} + \beta_{HCM} \,\nabla\times {\bf E}]\\[2pt]
{\bf B} = \muo\mu_{HCM}\, [{\bf H} + \beta_{HCM}\, \nabla\times {\bf H}]
\end{array}\right\}\,
\label{eq2}
\end{equation}
that presumably hold for the homogenized composite material (HCM). The exercise is well--founded only if the particles
of both component materials can be considered to be electrically small \c{LOCM}.

The Bruggeman approach for homogenization was initiated for isotropic mixtures of isotropic dielectric materials, 
but has been subsequently extended to far more complex situations \c{Mbook,Mackay}. The general formulation
of the approach is as follows: Suppose the composite material has been homogenized, and it obeys \r{eq2}. Disperse in
it, homogeneously and randomly, a small number density of particles of both types of component materials in the volumetric ratio $f_a:f_b$;
and then homogenize. The properties of the HCM could not have altered in consequence.

All particles of type $p$, $(p = a,b)$, are identical, and are equivalent to electric and magnetic dipole moments, ${\bf p}_p$ and ${\bf m}_p$,
when immersed in the HCM.
The standard interpretation of the  Bruggeman approach then requires the solution of the following two equations \c{WLM97}:
\begin{equation}
\left.\begin{array}{l}
f_a\,{\bf p}_a + f_b\,{\bf p}_b = {\bf 0}\\[2pt]
f_a\,{\bf m}_a + f_b\,{\bf m}_b = {\bf 0}
\end{array}\right\}\,.
\label{eq3}
\end{equation}
In the present
context, Kampia and Lakhtakia \c{KL} solved \r{eq3} for $\eps_{HCM}$, $\mu_{HCM}$ and $\beta_{HCM}$.

An alternative interpretation is that the dispersal of particles of component material $p$ is equivalent to the creation of
excess polarization and excess magnetization, ${\bf P}_p$ and ${\bf M}_p$, $(p = a,b)$, in the HCM. But the total excess polarization
and magnetization must be null--valued. Then, the two equations
\begin{equation}
\left.\begin{array}{l}
{\bf P}_a + {\bf P}_b = {\bf 0}\\[2pt]
{\bf M}_a +{\bf M}_b = {\bf 0}
\end{array}\right\}\,
\label{eq4}
\end{equation}
could be solved to determine $\eps_{HCM}$, $\mu_{HCM}$ and $\beta_{HCM}$. Although \r{eq4}
were stated by Kampia and Lakhtakia \c{KL}, these equations
were not solved by them; indeed,
expressions for ${\bf P}_p$ and ${\bf M}_p$ were not even provided by them. However, Shanker \c{Shan}
did present expressions for  ${\bf P}_p$ and ${\bf M}_p$, and then solved \r{eq4}.

\section{Numerical Results}
We decided to compare the implementations of \r{eq3} and \r{eq4}. All particles of both component materials
were treated as spheres of radius $R$. Expressions for the polarizability densities (relating electric and magnetic dipole
moments to exciting electric and magnetic fields) and polarization densities (relating excess polarization and excess magnetization 
to electric and magnetic fields) were obtained from Shanker's paper \c{Shan}.

Computed values of $\eps_{HCM}$, $\mu_{HCM}$ and $\beta_{HCM}$ as functions
of $f_b$ are shown in Figures 1 and 2  for
$\ko R \to 0$ and $\ko R=0.2$, where $\ko=\omega (\epso\muo)^{1/2}$ is the free--space wavenumber. The constitutive
properties of the component materials for the two figures are the same as chosen by Shanker \c{Shan}.

Quite clearly, Figures 1 and 2 show that the incorporation of the finite size of the particles gives rise to a dissipative HCM,
even when both component materials are nondissipative. This conclusion is true whenever a nonzero length--scale is considered
in a homogenization approach~---~whether as the particle size \c{Shan,D89}, or a correlation length for particle--distribution
statistics \c{Mackay}, or both \c{M04}. The incorporation of the length scale appears to account, in some manner, for the scattering loss.

More importantly, whether the length scale is neglected (Fig. 1) or considered (Fig. 2),  estimates of $\eps_{HCM}$, $\mu_{HCM}$ and $\beta_{HCM}$
from \r{eq3} and \r{eq4} do not coincide. There seems to be a basic difference between \r{eq3} and \r{eq4},
which persists even when $\mu_{a,b}=1$, $\beta_{a,b}=0$ and $R\to 0$. An explanation of this difference, in that simple
context for the sake of clarity,  is provided
in the next section.

\section{Explanation}

\subsection{Preliminaries} 
\label{prerem}
We begin with the derivation of an important equation.
Let all space be occupied by a homogeneous dielectric material with relative permittivity $\eps_{h}$ at the frequency of
interest; thus, its relevant frequency--domain constitutive relation is
\begin{equation}
\#D =\epso \eps_h\,\#E\,.
\end{equation}
Suppose that an electrically small sphere made of a dielectric material with relative permittivity $\eps_i$ were to
be introduced. This particle would act as an electric dipole moment
\begin{equation}
\#p = v\epso\,\alpha_{i/h}\,\tilde{\#e}\,,
\end{equation}
where $v$ is the volume of the particle, $\tilde{\#e}$ is the electric field at the location of the
particle if the particle were to be removed and the resulting hole filled with the host material, and the product of $\epso$ and
$\alpha_{i/h}$
is the polarizability density of the particle embedded in the specific host material. The exact expression of $\alpha_{i/h}$
does not matter for our purpose here \c{Lak92}; but we note that it is independent of $R$ for the Bruggeman approach,
and dependent on $R$ for the extended Bruggeman approach \c{PLS}.

Let  many identical particles be randomly dispersed in the host material, such that their number density $N$ is macroscopically
uniform. Then, the particles can be replaced by an {\em excess\/} polarization
\begin{equation}
\label{eqq4}
\#P = Nv\epso\,\alpha_{i/h}\,\tilde{\#E}\,,
\end{equation}
where
\begin{equation}
\label{eqq5}
\tilde{\#E} = \#E + \#P/3\epso\eps_h
\end{equation}
is the {\em local} electric field \c{Lak92}. The qualifier {\em excess} is used here because this $\#P$ is in addition to the
polarization $\epso (\eps_h-1)\,\#E$ that indicates the presence of the host material.

By virtue of \r{eqq4} and \r{eqq5}, the excess polarization
\begin{equation}
\#P = \epso\,\frac{f\alpha_{i/h}}{1-f\alpha_{i/h}/3\eps_h}\,\#E\,,
\label{exP}
\end{equation}
where $f=Nv$ is the volumetric fraction of the particulate material. Hence,
the constitutive relation of the HCM is
\begin{eqnarray}
\nonumber
\#D &=& \epso  \eps_h\#E +\#P
\\
\nonumber
&=&\epso\left[\eps_h+ \frac{f\alpha_{i/h}}{1-f\alpha_{i/h}/3\eps_h}\right]\#E
\\
&=&\epso\eps_{HCM}\,\#E\,,
\end{eqnarray}
so that
\begin{equation}
\eps_{HCM}= \eps_h+ \frac{f\alpha_{i/h}}{1-f\alpha_{i/h}/3\eps_h}
\label{basic}
\end{equation}
is the estimated relative permittivity of the HCM at the frequency of interest. The first rigorous derivation of
the foregoing equation can be attributed to Fax\'en \c{Fax}.

Parenthetically,
a Maxwell Garnett formula for $\eps_{HCM}$ can be derived by setting $\eps_h=\eps_a$ and $\eps_i=\eps_b$ in \r{basic},
which is quite appropriate if $f_b<f_a$;
otherwise, the choice $\left\{\eps_h=\eps_b,\,\eps_i=\eps_a\right\}$  should be made.
These two Maxwell Garnett estimates also
constitute the so--called Hashin--Shtrikman bounds on $\eps_{HCM}$ \c{HS}.

\subsection{Standard Interpretation of the Bruggeman approach: Eq. \r{eq3}}
\label{Brf}
As stated in Section 2,
let us imagine that the composite material has already been homogenized.
 Into this HCM, let spherical particles of both component materials be randomly dispersed. The combined volumetric fraction of the
particles introduced into the HCM is $f <<1$, with $ff_a$ and $ff_b$ being the respective volumetric fractions of the two
component materials in the particles. 
Hence,
\begin{equation}
\alpha_{i/h}= f_a\,\alpha_{a/HCM} + f_b\,\alpha_{b/HCM}
\end{equation}
is the  polarizability density of a {\em material--averaged\/} particle embedded in a material with $\eps_h=\eps_{HCM}$.
Equation \r{exP} then yields
\begin{equation}
\#P = \epso\,\frac{f\,( f_a\,\alpha_{a/HCM} + f_b\,\alpha_{b/HCM})}{1-f\,( f_a\,\alpha_{a/HCM} + f_b\,\alpha_{b/HCM})/3\eps_{HCM}}\,\#E\,,
\label{exPBr}
\end{equation}
for the excess polarization.

But the introduction of the  material--averaged particles must not change the HCM's constitutive
properties, as the relative proportion of the component materials remains unchanged; accordingly, the excess polarization of
\r{exPBr} is null--valued, and
 the solution of the equation
\begin{equation}
\label{Brugg}
0=f_a\,\alpha_{a/HCM} + f_b\,\alpha_{b/HCM}
\end{equation}
yields an estimate of $\eps_{HCM}$. Thus the standard interpretation of the Bruggeman approach leading
to \r{eq3} is as the {\em average polarizability density\/} approach.

\subsection{Shanker's Interpretation of the Bruggeman approach: Eq. \r{eq4}}

Once again, suppose that the composite material has been homogenized
into a HCM with relative permittivity $\eps_{HCM}$. Suppose, next, that particles of materials $a$ and $b$ are 
randomly dispersed the HCM and that their
respective volumetric fractions in the new composite material are $f_a$ and $f_b$. Following
Shanker \c{Shan}, we find that the excess polarizations due to the two types of 
particles add up to
\begin{equation}
\#P=\epso \left[\frac{f_a\alpha_{a/HCM}}{1-f_a\alpha_{a/HCM}/3\eps_{HCM}} +
\frac{f_b\alpha_{b/HCM}}{1-f_b\alpha_{b/HCM}/3\eps_{HCM}}\right]\#E
\,,
 \label{aNEP}
  \end{equation}
by virtue of \r{exP}. 

The introduction of the particles into the HCM amounts simply to the 
complete replacement of the HCM
by itself; hence, \r{basic} leads to
\begin{equation}
\eps_{HCM} = \eps_{HCM} + \frac{f_a\alpha_{a/HCM}}{1-f_a\alpha_{a/HCM}/3\eps_{HCM}} +
\frac{f_b\alpha_{b/HCM}}{1-f_b\alpha_{b/HCM}/3\eps_{HCM}}\,,
\end{equation}
which yields  the formula
 \begin{equation}
 \label{NEP}
 0= \frac{f_a\alpha_{a/HCM}}{1-f_a\alpha_{a/HCM}/3\eps_{HCM}} +
  \frac{f_b\alpha_{b/HCM}}{1-f_b\alpha_{b/HCM}/3\eps_{HCM}}
 \end{equation}
 for an estimate of $\eps_{HCM}$. Thus Shanker's interpretation of the Bruggeman approach leading
to \r{eq4} is as the {\em null excess polarization\/} approach.

\subsection{Comparison of the Two Interpretations}

Equation \r{aNEP} differs from \r{exPBr} in a very significant way: Whereas particles of
the two component materials were amalgamated into material--averaged particles whose polarizability density was used to
estimate the excess polarization as per \r{exPBr}, material--averaging was not done for \r{aNEP}; instead, particles of both materials
were kept apart and two separate contributions were made to the estimate \r{aNEP} of the excess polarization. 

This difference
can be understood also in terms of the different treatments of the local field. For \r{exPBr}, the local field pertains to material--averaged
particles, which is quite reasonable. In contrast, \r{aNEP} contains two different local fields. The first local field pertains {\em only\/}
to particles of material $a$ embedded in the HCM, and leads to the first term in the sum on the right side of \r{aNEP}; while
the second local field pertains {\em only\/}
to particles of material $b$ embedded in the HCM, and leads to the second term in the sum on the right side of \r{aNEP}. Accordingly,
\r{aNEP} lacks rigor in comparison to \r{exPBr}, and the former can be considered simply as an empirical formula.

In closing, if ${\bf P}_a$ and ${\bf P}_b$ could somehow be separately estimated in Shanker's interpretation with the same local field,
the two interpretations could very possibly yield identical estimates of the constitutive parameters of the HCM.

\bigskip
\noindent {\bf Acknowledgement.} The first author appreciates several discussions with Dr. Bernhard Michel.

\newpage
\begin{center}
\begin{figure}[h]
\centering \psfull \epsfig{file=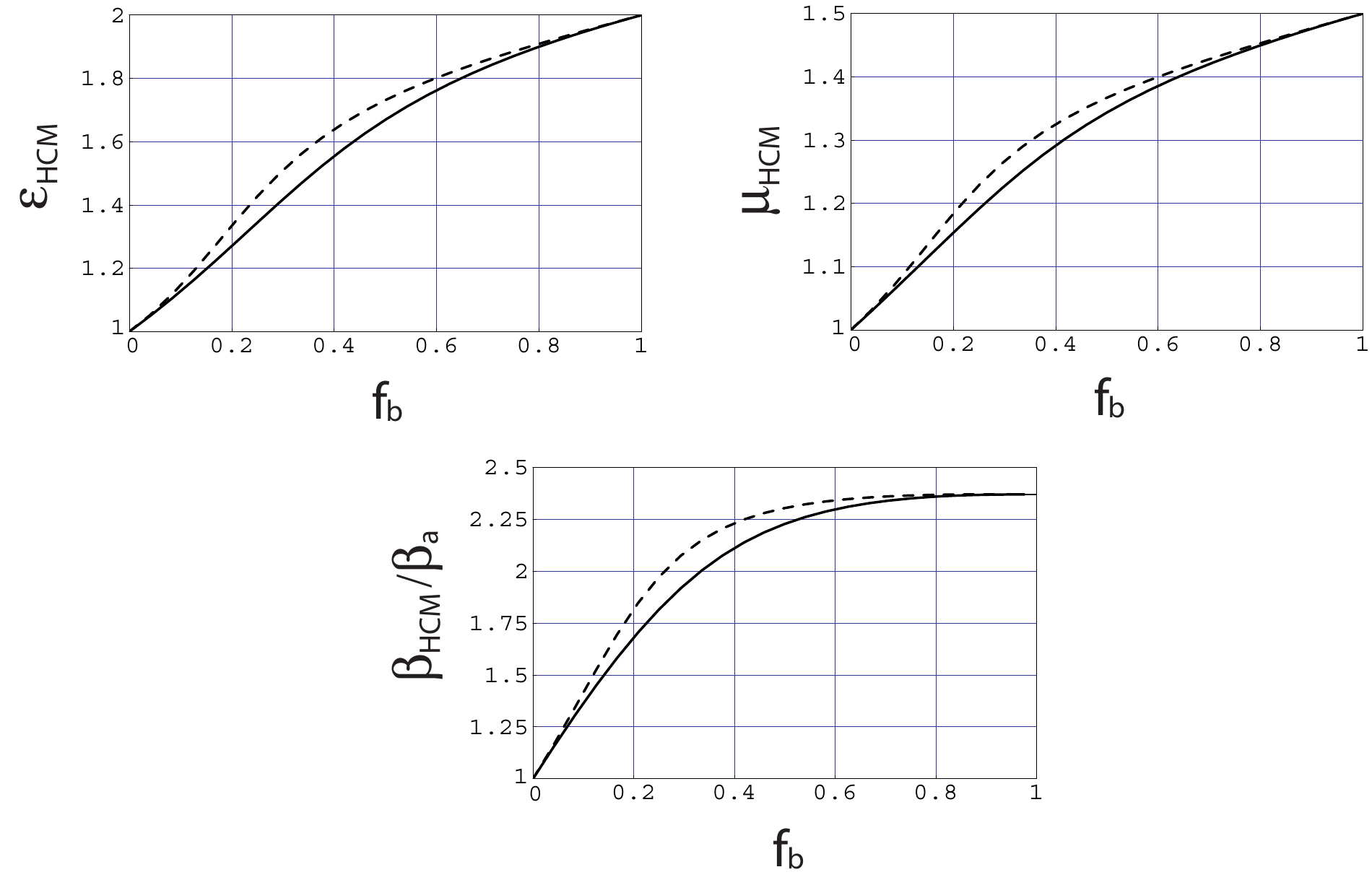,width=5in}
\caption{
Estimated values of $\eps_{HCM}$, $\mu_{HCM}$ and $\beta_{HCM}/\beta_a$ as functions of $f_b$, when
$\eps_a=1$, $\mu_a=1$, $\beta_a=10^{-3}$~m, 
$\eps_b=2$, $\mu_b=1.5$, $\beta_b=2.37\times10^{-3}$~m, and $\ko R\to0$. Solid lines represent data
computed using \r{eq3}, while dashed lines join datapoints obtained using \r{eq4}. }
\end{figure}
\end{center}

\newpage
\begin{center}
\begin{figure}[t]
\centering \psfull \epsfig{file=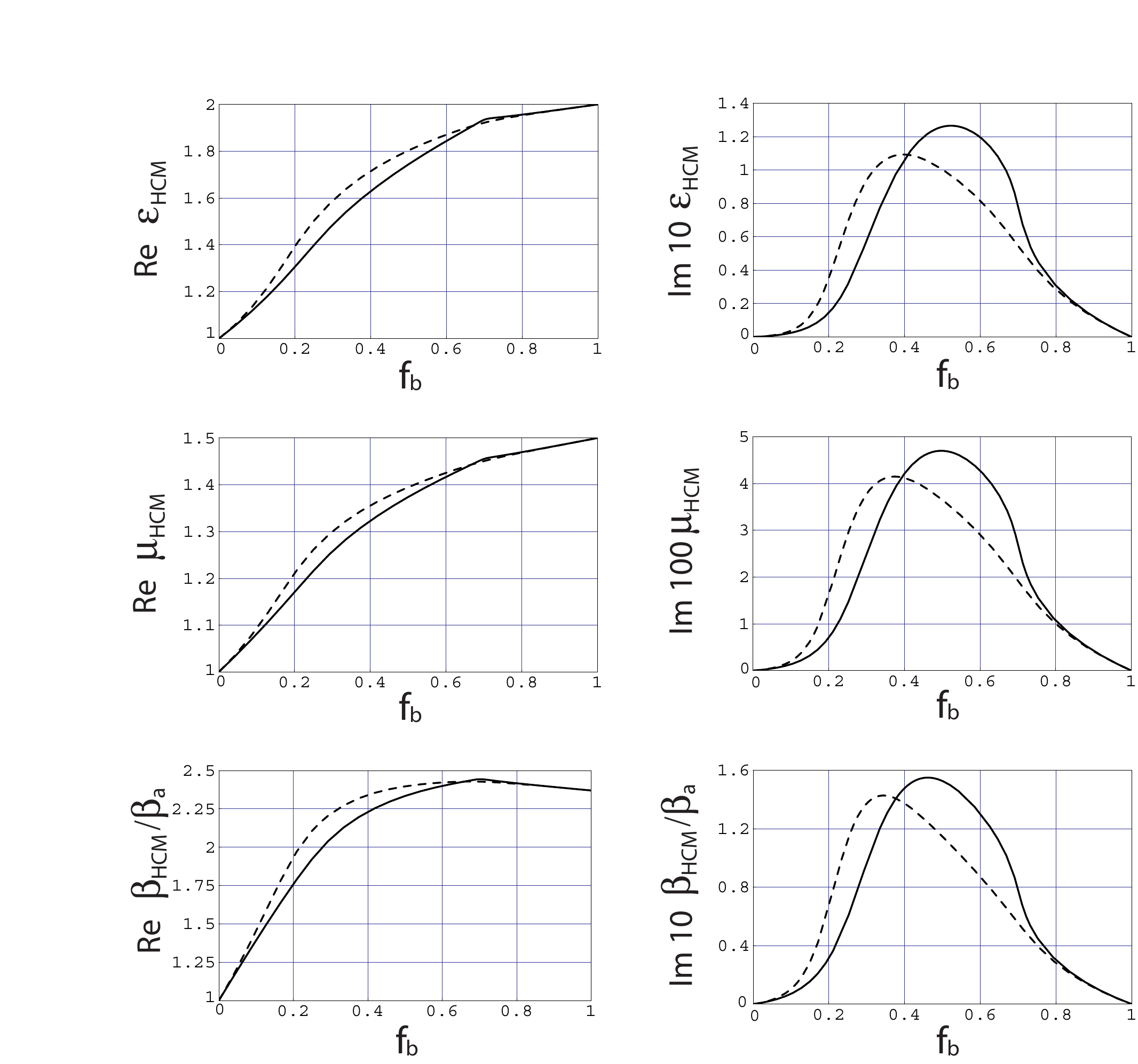,width=5in}
\caption{
Same as Figure 1, but for $\ko R=0.2$. 
}
\end{figure}
\end{center}

\end{document}